\documentclass[10pt,a4paper]{article}
\usepackage{amsmath,latexsym,amssymb,amsfonts}
\usepackage[dvips]{graphicx}
\usepackage{bm}

\begin{document}

\title{\textbf{Bubble collision with gravitation}}
\author{\textsc{Dong-il Hwang}$^{a,c}$\footnote{dongil.j.hwang@gmail.com},\;\; \textsc{Bum-Hoon Lee}$^{b,c}$\footnote{bhl@sogang.ac.kr},\;\; \textsc{Wonwoo Lee}$^{c}$\footnote{warrior@sogang.ac.kr},\\ and\;\; \textsc{Dong-han Yeom}$^{c,d}$\footnote{innocent.yeom@gmail.com}\\
\textit{$^{a}$\small{Department of Physics, KAIST, Daejeon 305-701, Republic of Korea}}\\
\textit{$^{b}$\small{Department of Physics, Sogang University, Seoul 121-742, Republic of Korea}}\\
\textit{$^{c}$\small{Center for Quantum Spacetime, Sogang University, Seoul 121-742, Republic of Korea}}\\
\textit{$^{d}$\small{Research Institute for Basic Science, Sogang University, Seoul 121-742, Republic of Korea}}}
\maketitle

\begin{abstract}
In this paper, we study vacuum bubble collisions with various potentials including gravitation, assuming spherical, planar, and hyperbolic symmetry. We use numerical calculations from double-null formalism. Spherical symmetry can mimic the formation of a black hole via multiple bubble collisions. Planar and especially hyperbolic symmetry describes two bubble collisions. We study both cases, when two true vacuum regions have the same field value or different field values, by varying tensions. For the latter case, we also test symmetric and asymmetric bubble collisions, and see details of causal structures. If the colliding energy is sufficient, then the vacuum can be destabilized, and it is also demonstrated. This double-null formalism can be a complementary approach in the context of bubble collisions.
\end{abstract}

\newpage

\tableofcontents

\newpage

\section{Introduction}

Nucleation and subsequent evolution of vacuum bubbles are very interesting issues in gravity, field theory, and cosmology. We know that our universe experiences accelerated expansion and this state may be described by a positive cosmological constant. In addition, we know that our universe had experienced a stage of inflation. The primordial inflation can be described by the existence of a false vacuum in a potential of a field and this necessarily requires a phase transition of the field from the false vacuum to the current true vacuum. In this context, the nucleation and evolution of vacuum bubbles are quite natural to introduce. In addition, such a study of vacuum bubbles can be interesting and important.

The primordial inflation of the current universe will probably be described by a potential with some fields that causes inflation \cite{Guth:1980zm}. One of the important points of consensus is that the phase transition that finishes the primordial inflation seems to be slow-rolling, not the first order phase transition \cite{Linde:1981mu}. In other words, it is difficult to explain the ending of inflation by the nucleation and percolation of true vacuum bubbles, since the inflating universe quickly expands, so that bubbles in general cannot percolate sufficiently \cite{Guth:1982pn}. Although we know this pessimistic information, the study of bubble nucleation and dynamics is still quite important. There are at least two main reasons why.
\begin{enumerate}
\item \textbf{Cosmic landscape and multiverse:} String theory requires moduli stabilization and this may be possible by flux compactification \cite{Kachru:2003aw}. The flux compactification can generate a huge number of different vacua \cite{Bousso:2000xa}, the so-called cosmic landscape \cite{Susskind:2003kw}. If eternal inflation happens, then all vacua can be populated by first order phase transitions (quantum tunneling) and each bubble will form a pocket universe. The totality of pocket universes in the eternally inflating background is called the multiverse. In the multiverse and cosmic landscape, study of the nucleation and subsequent dynamics of vacuum bubbles is illuminated again.
\item \textbf{In the context of thermal inflation:} After the primordial inflation ended, some unnecessary massive particles can be overproduced \cite{Lyth:1995ka}. Such massive particles should disappear, and this may be possible by introducing thermal inflation. In many phenomenological models of thermal inflation, it will be ended by the first order phase transition, not the second order phase transition \cite{Easther:2008sx}. Therefore, our universe may have some signs of the first order phase transition, bubble nucleation, and bubble percolation.
\end{enumerate}

The bubble nucleating process including gravitation was first studied by Coleman and DeLuccia \cite{Coleman:1980aw}. If we include non-minimal coupling effects that can be motivated from string theory, then possible vacuum transitions can be complicated \cite{Lee:2006vka}. The subsequent evolution of the bubble including gravitation can be approximated by the thin-wall approximation \cite{Blau:1986cw}. However, the thin-wall approximation itself is an approximation and hence it may lose dynamics at the transition region. The related study beyond the thin-wall approximation was studied by numerical method \cite{Hansen:2009kn}\cite{Hwang:2010gc}.

In this paper, we focus on bubble collision issues. This issue has a long history. First, without considering gravitation, there were some studies using analytic methods or numerical methods \cite{Hawking:1982ga}. One of the interesting points of colliding bubbles is that the colliding bubble may induce a vacuum transition \cite{Johnson:2010bn} or a production of particles \cite{Zhang:2010qg} using the colliding energy. Second, including gravitation, it is highly non-trivial to study the dynamics. We may use the thin-wall approximation \cite{Wu:1984eda}\cite{Moss:1994iq}. In the study of Freivogel, Horowitz, and Shenker \cite{Freivogel:2007fx} (for more advanced review, see \cite{Chang:2007eq}), they discussed two colliding true vacuum bubbles in the de Sitter background: one is flat and the other is anti de Sitter. The analysis in itself is very concise and important, but this may not be able to describe the dynamics of fields on the colliding walls. For example, this cannot describe the vacuum transition behaviors. Hence, we may need further numerical studies.

There were some numerical studies of bubble collisions with gravitation. Very recently, Johnson, Peiris, and Lehner \cite{Johnson:2011wt} succeeded in studying bubble collisions with gravitation beyond the thin-wall approximation. They assumed hyperbolic symmetry from the Birkhoff-like theorem of colliding bubbles and assumed initial data from the Coleman-DeLuccia type solutions. They could solve Einstein and field equations numerically and observe and report on symmetric/asymmetric bubble collisions and vacuum transitions.

In this paper, our study can be an alternative approach to the problem using the double-null formalism \cite{Hamade:1995ce}\cite{Hong:2008mw} (note that, \cite{BlancoPillado:2003hq} studied $5$-dimensional anti de Sitter bubble collisions for different purposes). We can list possible differences with the previous studies:
\begin{enumerate}
\item \textbf{Symmetry}: When two bubbles are colliding and there are no other perturbations that disturb bubble collisions, then we can assume the hyperbolic symmetry. However, if many bubbles collide and the background is very complicated, then the Birkhoff-like theorem can be highly non-trivial. For the multi-bubble colliding case \cite{Moss:1994iq}, the spherical symmetry can be a simplified model; also, it may be useful to try to extend the symmetry to be not only hyperbolic, but also planar.

\item \textbf{Beyond the Coleman-DeLuccia conditions}: At once we fix a certain symmetry (spherical, hyperbolic, or planar), we can assign various initial conditions for the field within the symmetry. These initial conditions can come from the Coleman-DeLuccia solution, but it is not necessary in general\footnote{For a general spherical bubble, it may be difficult to give an exact relation between a spherical slice and a hyperbolic slice. However, we can justify the hyperbolic `approximation' at least for the vicinity of the colliding walls when each walls become sufficiently large. Since the lapse function is approximately $\sim 1 - 2M/r + \mathcal{O}(r^{2})$, around the large wall of a spherical bubble, one can approximate de Sitter, flat, or anti de Sitter. In these limits, we can justify hyperbolic slices from spherical slices. The planar or hyperbolic symmetry can be the best trial for general spherical bubble collisions beyond the Coleman-DeLuccia conditions: we can try to see qualitative results and we can choose a certain symmetry by a practical way. The authors thank to a referee to point this out.}. The Coleman-DeLuccia solution assumes $O(4)$ symmetry. Therefore, their solutions are de Sitter, Minkowski, or anti de Sitter. Then the initial tension and the initial size of the bubble are already fixed. However, in general, bubbles can be massive and hence the tension and size can be free parameters \cite{Blau:1986cw}, although the probability of the bubble can be reduced. It is reasonable that if a massive bubble is nucleated in the de Sitter background, then the causal future of the massive bubble can see the effect of the mass, while the causal past may not notice such behavior \cite{Hansen:2009kn}\cite{Hwang:2010gc}. In the double-null coordinates, it is very easy to implement this kind of massive thick-wall conditions.

\item \textbf{Causal structure}: The double-null formalism does not have coordinate singularities in general, since it does not use the coordinate time. In addition, the double-null coordinate preserves causality. Therefore, it is easy to draw causal structures for complicated cases.
\end{enumerate}

This paper is organized as follows. In Section~\ref{sec:mod}, we discuss our model for bubble collisions in the double-null formalism with spherical, planar, and hyperbolic symmetry. In Section~\ref{sec:var}, we show various phenomena of bubble collisions: bubble percolation, dynamics of bubble walls, and vacuum destabilization. Finally, in Section~\ref{sec:dis}, we summarize our results.

\begin{figure}
\begin{center}
\includegraphics[scale=0.4]{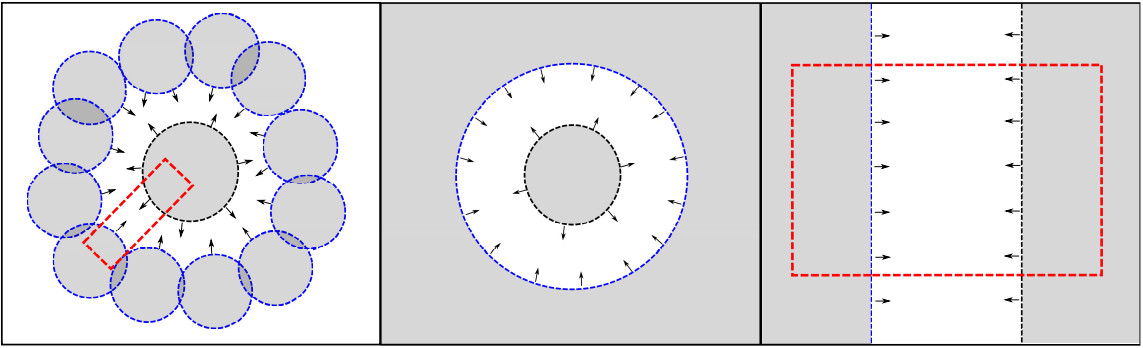}
\caption{\label{fig:bubbles}Left: General bubble percolation. Gray regions are true vacuum regions and white regions are false vacuum regions. Middle: Approximately, this can be described by an out-going true vacuum bubble and an in-going false vacuum bubble. Right: When two bubbles are colliding, we can magnify around the colliding shells, so that the region can be approximated by left-moving and right-moving walls.}
\end{center}
\end{figure}

\section{\label{sec:mod}Model for bubble collisions}

In this paper, we study a gravitational system in Einstein gravity with a scalar field that is governed by a potential. Then the action $S$ becomes:
\begin{eqnarray}
S = \int dx^{4} \sqrt{-g} \left[ \frac{1}{16\pi} R - \frac{1}{2}\nabla_{\mu} \phi \nabla^{\mu} \phi - V(\phi) \right],
\end{eqnarray}
where $R$ is the Ricci scalar, $\phi$ is the scalar field, and $V(\phi)$ is the potential of the scalar field.

Einstein equations and scalar field equations are as follows:
\begin{eqnarray}\label{eq:Einstein}
G_{\mu\nu} &=& 8 \pi T^{\phi}_{\mu\nu} \\
\label{eq:T_Phi}
T^{\phi}_{\mu\nu} &=& \phi_{;\mu}{\phi}_{;\nu} - \frac{1}{2}\phi_{;\rho}\phi_{;\sigma}g^{\rho\sigma}g_{\mu\nu} -V(\phi) g_{\mu\nu} \\
\label{eq:Phi}0 &=& \phi_{;\mu\nu}g^{\mu\nu}- \frac{dV}{d\phi}
\end{eqnarray}

In this paper, we will choose one of the following metric ansatz of double-null coordinate:
\begin{enumerate}
\item Spherical symmetry: $dS^{2} = d\theta^{2} + \sin^{2} \theta d\varphi^{2}$
\begin{eqnarray}
ds^{2} = -\alpha_{\mathrm{s}}^{2}(u,v) du dv + r_{\mathrm{s}}^{2}(u,v) dS^{2}
\end{eqnarray}
\item Planar symmetry: $dR^{2} = dx^{2} + dy^{2}$
\begin{eqnarray}
ds^{2} = -\alpha_{\mathrm{p}}^{2}(u,v) du dv + r_{\mathrm{p}}^{2}(u,v) dR^{2}
\end{eqnarray}
\item Hyperbolic symmetry: $dH^{2} = d\chi^{2} + \sinh^{2} \chi d\varphi^{2}$
\begin{eqnarray}
ds^{2} = -\alpha_{\mathrm{h}}^{2}(u,v) du dv + r_{\mathrm{h}}^{2}(u,v) dH^{2}
\end{eqnarray}
\end{enumerate}
Note that $0 \leq \theta \leq \pi$, $0 \leq \varphi < 2\pi$, $-\infty < x,y < \infty$, and $0 \leq \chi < \infty$.

Figure~\ref{fig:bubbles} shows general motivations of each symmetries. When vacuum bubbles are colliding and percolated, there can be a certain true vacuum bubble that is surrounded by a false vacuum region, where the outside of the false vacuum region is already transited to the true vacuum (left). We may be able to approximate such a situation by an in-going false vacuum bubble and an out-going true vacuum bubble with spherical symmetry (middle). In addition, if we focus on the collision of two vacuum bubbles (red square in left), the colliding region locally looks like two walls colliding (right). Such walls can be approximated by the planar symmetry or the hyperbolic symmetry. 

\subsection{Comments on mass functions}

To give the proper initial conditions, first we have to understand the value of metric ansatz in the static limit. For spherical and hyperbolic cases, we can impose the Birkhoff theorem and obtain the concepts of the local mass functions. We know that the general solutions for $V(\phi)=V_{0}$ will look like
\begin{eqnarray}
ds^{2} = - N(r)^{2} dt^{2} + \frac{{dr}^{2}}{N(r)^{2}} + r^{2} d\Omega_{\kappa}^{2},
\end{eqnarray}
where $\kappa = +1$, $0$, or $-1$,
\begin{eqnarray}
N^{2} = \kappa - \frac{2M}{r} - \frac{8\pi V_{0} r^{2}}{3},
\end{eqnarray}
and $d\Omega_{+1}=dS$, $d\Omega_{0}=dR$, or $d\Omega_{-1}=dH$, respectively \cite{Freivogel:2007fx}.

\begin{enumerate}
\item The solution in the spherical symmetry is
\begin{eqnarray}
ds^{2} = - N_{\mathrm{s}}(r_{\mathrm{s}})^{2} dt_{\mathrm{s}}^{2} + \frac{dr_{\mathrm{s}}^{2}}{N_{\mathrm{s}}(r_{\mathrm{s}})^{2}} + r_{\mathrm{s}}^{2} dS^{2},
\end{eqnarray}
where
\begin{eqnarray}
N_{\mathrm{s}}^{2} = 1 - \frac{2M}{r_{\mathrm{s}}} - \frac{8\pi V r_{\mathrm{s}}^{2}}{3}.
\end{eqnarray}
Here, $r_{\mathrm{s}}$ is the space-like parameter and $t_{\mathrm{s}}$ is the time-like parameter.
We define coordinate transformation:
\begin{eqnarray}
dr_{\mathrm{s}} &=& {r_{\mathrm{s}}}_{,u}du + {r_{\mathrm{s}}}_{,v}dv,\\
dt_{\mathrm{s}} &=& \frac{\alpha_{\mathrm{s}}^{2}}{4} \left( -\frac{dv}{{r_{\mathrm{s}}}_{,u}} + \frac{du}{{r_{\mathrm{s}}}_{,v}} \right),
\end{eqnarray}
and obtain the double-null metric $ds^{2} = -\alpha_{\mathrm{s}}^{2}(u,v) du dv + r_{\mathrm{s}}^{2}(u,v) dS^{2}$. Note that we can choose the in-going direction ${r_{\mathrm{s}}}_{,u}<0$ and the out-going direction ${r_{\mathrm{s}}}_{,v}>0$.
Thus, we can show
\begin{eqnarray}
N_{\mathrm{s}}(r_{\mathrm{s}})^{2} = - \frac{4{r_{\mathrm{s}}}_{,u}{r_{\mathrm{s}}}_{,v}}{\alpha^{2}}.
\end{eqnarray}
Therefore, we can identify that the Misner-Sharp mass function is
\begin{eqnarray}
m_{\mathrm{s}}(u,v) = \frac{r_{\mathrm{s}}}{2} \left( 1 + \frac{4 {r_{\mathrm{s}}}_{,u} {r_{\mathrm{s}}}_{,v}}{\alpha_{\mathrm{s}}^{2}} - \frac{8\pi V}{3} {r_{\mathrm{s}}}^{2} \right).
\end{eqnarray}

\item The solution in the hyperbolic symmetry will look like
\begin{eqnarray}
ds^{2} = - N_{\mathrm{h}}(r_{\mathrm{h}})^{2} dt_{\mathrm{h}}^{2} + \frac{dr_{\mathrm{h}}^{2}}{N_{\mathrm{h}}(r_{\mathrm{h}})^{2}} + r_{\mathrm{h}}^{2} dH^{2},
\end{eqnarray}
where
\begin{eqnarray}
N_{\mathrm{h}}^{2} = - \left( 1 + \frac{2M}{r_{\mathrm{h}}} + \frac{8\pi V r_{\mathrm{h}}^{2}}{3} \right)
\end{eqnarray}
from the Birkhoff theorem and $N_{\mathrm{h}}^{2}$ is chosen to be negative. Therefore, now $t_{\mathrm{h}}$ is the space-like parameter and $r_{\mathrm{h}}$ is the time-like parameter.
We define coordinate transformation:
\begin{eqnarray}
dr_{\mathrm{h}} &=& {r_{\mathrm{h}}}_{,u}du + {r_{\mathrm{h}}}_{,v}dv,\\
dt_{\mathrm{h}} &=& \frac{\alpha_{\mathrm{h}}^{2}}{4} \left( - \frac{dv}{{r_{\mathrm{h}}}_{,u}} + \frac{du}{{r_{\mathrm{h}}}_{,v}} \right),
\end{eqnarray}
and obtain the double-null metric $ds^{2} = -\alpha_{\mathrm{h}}^{2}(u,v) du dv + r_{\mathrm{h}}^{2}(u,v) dH^{2}$. Note that we should choose both left-moving and right-moving null directions ${r_{\mathrm{h}}}_{,u}>0$ and ${r_{\mathrm{h}}}_{,v}>0$.
Thus, we can show
\begin{eqnarray}
N_{\mathrm{h}}^{2} = -\frac{4{r_{\mathrm{h}}}_{,u}{r_{\mathrm{h}}}_{,v}}{\alpha_{\mathrm{h}}^{2}}.
\end{eqnarray}
Therefore, we can identify that the mass function in the double-null coordinate by
\begin{eqnarray}
m_{\mathrm{h}}(u,v) = -\frac{r_{\mathrm{h}}}{2} \left( 1 - \frac{4 {r_{\mathrm{h}}}_{,u} {r_{\mathrm{h}}}_{,v}}{\alpha_{\mathrm{h}}^{2}} + \frac{8\pi V}{3} r_{\mathrm{h}}^{2} \right).
\end{eqnarray}
Note that usual black hole type solutions in Minkowski vacuum will happen for $m_{\mathrm{h}} < 0$ limit, and hence, in this paper, we are interested in $m_{\mathrm{h}} \leq 0$.
\end{enumerate}

\subsection{\label{sec:imp}Implementation of double-null formalism}

From now, if there is no confusion, then we will omit the subscripts $\mathrm{s}$, $\mathrm{p}$, or $\mathrm{h}$.

Define
\begin{eqnarray}
\sqrt{4\pi}\phi \equiv S
\end{eqnarray}
and use conventions \cite{Hong:2008mw}
\begin{eqnarray}\label{eq:conventions}
h \equiv \frac{\alpha_{,u}}{\alpha},\quad d \equiv \frac{\alpha_{,v}}{\alpha},\quad f \equiv r_{,u},\quad g \equiv r_{,v},\quad W \equiv S_{,u},\quad Z \equiv S_{,v}.
\end{eqnarray}

Einstein tensor components are
\begin{eqnarray}
\label{eq:Guu}G_{uu} &=& -\frac{2}{r} \left(f_{,u}-2fh \right),\\
\label{eq:Guv}G_{uv} &=& \frac{1}{2r^{2}} \left( 4 rf_{,v} + \kappa \alpha^{2} + 4fg \right),\\
\label{eq:Gvv}G_{vv} &=& -\frac{2}{r} \left(g_{,v}-2gd \right),\\
\label{eq:Gthth}G_{aa} &=& -4\frac{r^{2}}{\alpha^{2}} \left(d_{,u}+\frac{f_{,v}}{r}\right),
\end{eqnarray}
where $\kappa=+1, 0, -1$ and $a=\theta, x, \chi$ for spherical, planar, hyperbolic cases, respectively.
Energy-momentum tensor components are
\begin{eqnarray}
\label{eq:TPhiuu}T^{\phi}_{uu} &=& \frac{1}{4 \pi} W^{2},\\
\label{eq:TPhiuv}T^{\phi}_{uv} &=& \frac{\alpha^{2}}{2} V(S),\\
\label{eq:TPhivv}T^{\phi}_{vv} &=& \frac{1}{4 \pi} Z^{2},\\
\label{eq:TPhithth}T^{\phi}_{aa} &=& \frac{r^{2}}{2 \pi \alpha^{2}} WZ -r^{2}V(S).
\end{eqnarray}

Therefore, simulation equations are as follows:
\begin{eqnarray}
\label{eq:E1}f_{,u} &=& 2 f h - 4 \pi r T^{\phi}_{uu},\\
\label{eq:E2}g_{,v} &=& 2 g d - 4 \pi r T^{\phi}_{vv},\\
\label{eq:E3}f_{,v}=g_{,u} &=& - \kappa \frac{\alpha^{2}}{4r} - \frac{fg}{r} + 4\pi r T^{\phi}_{uv},\\
\label{eq:E4}h_{,v}=d_{,u} &=& -\frac{2\pi \alpha^{2}}{r^{2}}T^{\phi}_{aa} - \frac{f_{,v}}{r}.
\end{eqnarray}
In addition, we include the scalar field equation:
\begin{eqnarray}
\label{eq:S}Z_{,u} = W_{,v} = - \frac{fZ}{r} - \frac{gW}{r} -\pi\alpha^{2} V'(S).
\end{eqnarray}

\subsection{\label{sec:ini}Initial conditions}

We need initial conditions for all functions ($\alpha, h, d, r, f, g, S, W, Z$) on the initial $u=u_{\mathrm{i}}$ and $v=v_{\mathrm{i}}$ surfaces, where we set $u_{\mathrm{i}}=v_{\mathrm{i}}=0$.

\paragraph{Spherical symmetry}

We have gauge freedom to choose the initial $r$ function. Although all constant $u$ and $v$ lines are null, there remains freedom to choose the distances between these null lines. Here, we choose $r(0,0)=r_{0}$, $f(u,0)=r_{u0}$, and $g(0,v)=r_{v0}$, where $r_{u0}<0$ and $r_{v0}>0$ such that the radial function for an in-going observer decreases and that for an out-going observer increases.

It is convenient to choose $r_{u0}=-1/2$ and $r_{v0}=1/2$; we choose that the mass function on $u_{\mathrm{i}}=v_{\mathrm{i}}=0$ vanish, since we can think that the initial point is in the causal past of the bubbles. Hence, to specify a false vacuum background, for given $r(0,0)=r_{0}$ and $S(0,0)=S_{\mathrm{f}}$, and if the field is at the local minimum, then
\begin{eqnarray}
\alpha(0,0) = \left( 1 - \frac{8 \pi V(S_{\mathrm{f}}) r_{0}^{2}}{3} \right)^{-1/2}.
\end{eqnarray}

\paragraph{Planar symmetry}

We first choose $r(0,0)=r_{0}$, $f(0,0)=0$, $g(0,0)=0$, $\alpha(u,0)=1$, and $\alpha(0,v)=1$, for simplicity. Therefore, $h(u,0)=d(0,v)=0$.

\paragraph{Hyperbolic symmetry}

We have gauge freedom to choose the initial $r$ function. Although all constant $u$ and $v$ lines are null, there remains freedom to choose the distances between these null lines. Here, we choose $r(0,0)=r_{0}$, $f(u,0)=r_{u0}$, and $g(0,v)=r_{v0}$, where $r_{u0}>0$ and $r_{v0}>0$ such that the function $r$ increases for both of in-going and out-going observers.

It is convenient to choose $r_{u0}=1/2$ and $r_{v0}=1/2$; we choose that the mass function on $u_{\mathrm{i}}=v_{\mathrm{i}}=0$ vanish. Hence, to specify a pure de Sitter background, for given $r(0,0)=r_{0}$, $S(0,0)=S_{\mathrm{f}}$, and if the field is at the local minimum, then
\begin{eqnarray}
\alpha(0,0) = \left( 1 + \frac{8 \pi V(S_{\mathrm{f}}) r_{0}^{2}}{3} \right)^{-1/2}.
\end{eqnarray}
To compare previous studies, we can choose the initial mass function nonzero. Then,
\begin{eqnarray}
\alpha(0,0) = \left( 1 + \frac{8 \pi V(S_{\mathrm{f}}) r_{0}^{2}}{3} + \frac{2 m_{0}}{r_{0}} \right)^{-1/2}
\end{eqnarray}
with a free parameter $m_{0}$.

\paragraph{Assignments of initial conditions}

Now it is possible to assign all of initial conditions along initial $u=u_{\mathrm{i}}$ and $v=v_{\mathrm{i}}$ surfaces.

\begin{description}
\item[Initial $v=v_{\mathrm{i}}$ surface:]
We choose
\begin{eqnarray}
S(u,0) = \left\{ \begin{array}{ll}
S_{\mathrm{f}} & u < u_{\mathrm{shell}},\\
\left| S_{\mathrm{t}}-S_{\mathrm{f}} \right| G(u) + S_{\mathrm{t}} & u_{\mathrm{shell}} \leq u < u_{\mathrm{shell}}+\Delta u,\\
S_{\mathrm{t}} & u_{\mathrm{shell}}+\Delta u \leq u,
\end{array} \right.
\end{eqnarray}
where $G(u)$ is a pasting function which goes from $1$ to $0$ by a smooth way. We choose $G(u)$ by
\begin{eqnarray}
G(u) = 1 - \sin^{2} \left[\frac{\pi(u-u_{\mathrm{shell}})}{2\Delta u}\right].
\end{eqnarray}
Then, we know $W(u,0)=S_{,u}(u,0)$. If $\kappa=\pm1$, then $h(u,0)$ is given from Equation~(\ref{eq:E1}), since $f_{,u}=0$ along the in-going null surface. Then, using $h(u,0)$, we obtain $\alpha(u,0)$. If $\kappa=0$, then we obtain $f(u,0)$ from Equation~(\ref{eq:E1}).

We need more information to determine $d, g$, and $Z$ on the $v=0$ surface. We obtain $d$ from Equation~(\ref{eq:E4}), $g$ from Equation~(\ref{eq:E3}), and $Z$ from Equation~(\ref{eq:S}).

\item[Initial $u=u_{\mathrm{i}}$ surface:]
We choose
\begin{eqnarray}
S(0,v) = \left\{ \begin{array}{ll}
S_{\mathrm{f}} & v < v_{\mathrm{shell}},\\
\left| S_{\mathrm{f}} - S_{\mathrm{t}} \right| G(v) - S_{\mathrm{t}} & v_{\mathrm{shell}} \leq v < v_{\mathrm{shell}}+\Delta v,\\
S_{\mathrm{t}} & v_{\mathrm{shell}}+\Delta v \leq v,
\end{array} \right.
\end{eqnarray}
where
\begin{eqnarray}
G(v) = 1 - \sin^{2} \left[\frac{\pi(v-v_{\mathrm{shell}})}{2\Delta v}\right].
\end{eqnarray}

If $\kappa=\pm1$, then we obtain $d(0,v)$ from Equation~(\ref{eq:E2}), since $g_{,v}(0,v)=0$. If $\kappa=0$, then we obtain $g(0,v)$ from Equation~(\ref{eq:E2}). By integrating $d$ along $v$, we have $\alpha(0,v)$.

We need more information for $h, f,$ and $W$ on the $u=0$ surface. We obtain $h$ from Equation~(\ref{eq:E4}), $f$ from Equation~(\ref{eq:E3}), and $W$ from Equation~(\ref{eq:S}). This finishes the assignments of the initial conditions.
\end{description}

We used the second order Runge-Kutta method. The details of the numerical scheme and convergence/consistency checks are included in Appendix A and B.


\subsection{Simulation parameters}

Finally, we can specify all simulation parameters. In addition, we need information on potentials for detailed purposes.

We illustrate all simulation parameters.
\begin{enumerate}
\item Initial $r$ and Misner-Sharp mass functions: $r_{0}$, $m_{0}$
\item Thickness of the shells: $\Delta u$, $\Delta v$
\item Size of the shells: $u_{\mathrm{shell}}$, $v_{\mathrm{shell}}$
\item Field values: $S_{\mathrm{t}}$, $S_{\mathrm{f}}$
\item Potentials
\end{enumerate}

\begin{figure}
\begin{center}
\includegraphics[scale=0.6]{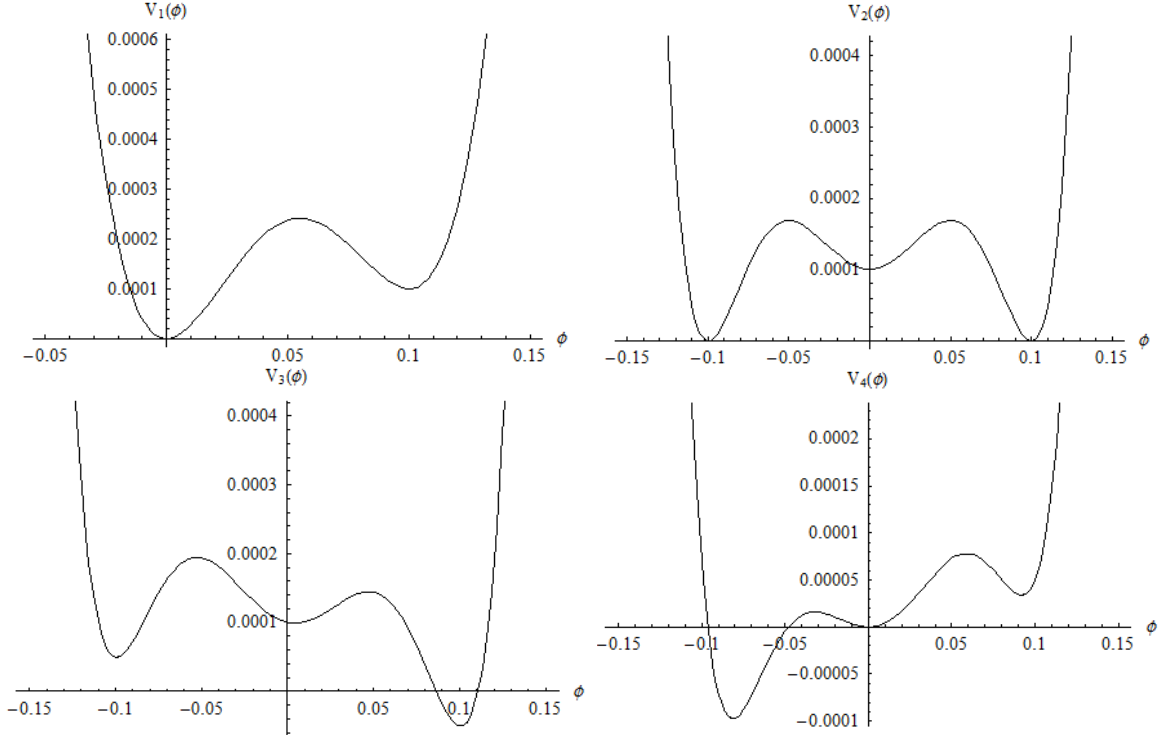}
\caption{\label{fig:pot}Potentials we used. $V_{\mathrm{f}}=0.0001$ and $\phi_{\mathrm{f}}$ or $\phi_{\mathrm{t}}$ are chosen to be $0.1$.}
\end{center}
\end{figure}

Especially, we need further comments on the potentials (Figure~\ref{fig:pot}).
\begin{enumerate}
\item To study bubble percolation, we use the potential
\begin{eqnarray}
V_{1} (\phi) = \frac{V_{\mathrm{f}}}{12 \beta \phi_{\mathrm{f}}} \int_{0}^{\phi} \frac{\bar{\phi}}{\phi_{\mathrm{f}}} \left( \frac{\bar{\phi}}{\phi_{\mathrm{f}}} - \frac{\beta+1}{2} \right) \left( \frac{\bar{\phi}}{\phi_{\mathrm{f}}} -1 \right) d \bar{\phi},
\end{eqnarray}
where $\beta = 0.1$ is a free parameter, $V_{\mathrm{f}}$ is the vacuum energy of the false vacuum, and $\phi_{\mathrm{f}}$ is the field value of the false vacuum.

\item To study bubble collision with different true vacua, we use the potential
\begin{eqnarray}
V_{2} (\phi) = \frac{V_{\mathrm{f}}}{C_{2}} \int_{\phi_{\mathrm{t}}}^{\phi} \left[ \frac{\bar{\phi}}{\phi_{\mathrm{t}}} \left( \frac{\bar{\phi}}{\phi_{\mathrm{t}}} - 1 \right) \left( \frac{\bar{\phi}}{\phi_{\mathrm{t}}} + 1 \right) \left( \frac{\bar{\phi}}{\phi_{\mathrm{t}}} - \frac{1}{2} \right) \left( \frac{\bar{\phi}}{\phi_{\mathrm{t}}} + \frac{1}{2} \right) \right] d \bar{\phi},
\end{eqnarray}
where $V_{\mathrm{f}}$ is the vacuum energy of the false vacuum and $C_{2}$ is a normalization constant with
\begin{eqnarray}
C_{2} = \int_{\phi_{\mathrm{t}}}^{0} \left[ \frac{\bar{\phi}}{\phi_{\mathrm{t}}} \left( \frac{\bar{\phi}}{\phi_{\mathrm{t}}} - 1 \right) \left( \frac{\bar{\phi}}{\phi_{\mathrm{t}}} + 1 \right) \left( \frac{\bar{\phi}}{\phi_{\mathrm{t}}} - \frac{1}{2} \right) \left( \frac{\bar{\phi}}{\phi_{\mathrm{t}}} + \frac{1}{2} \right) \right] d \bar{\phi},
\end{eqnarray}
and $\phi_{\mathrm{t}}$ is the field difference between the false vacuum and the true vacuum.

\item If we tilt the potential $V_{2}$ as follows:
\begin{eqnarray}
V_{3} (\phi) = V_{2}(\phi) - D_{3} \frac{\phi}{\phi_{\mathrm{t}}},
\end{eqnarray}
where $D_{3}/V_{\mathrm{f}} = 0.5$ is fixed, we then observe the asymmetric bubble collisions.

\item To study vacuum destabilization by bubble collisions, we use the potential:
\begin{eqnarray}
V_{4} (\phi) = \frac{V_{\mathrm{f}}}{C_{4}} \int_{\phi_{\mathrm{f}}}^{\phi} \left[ \frac{\bar{\phi}}{\phi_{\mathrm{f}}} \left( \frac{\bar{\phi}}{\phi_{\mathrm{f}}} - 1 \right) \left( \frac{\bar{\phi}}{\phi_{\mathrm{f}}} + 0.75 \right) \left( \frac{\bar{\phi}}{\phi_{\mathrm{f}}} - \frac{1}{2} \right) \left( \frac{\bar{\phi}}{\phi_{\mathrm{f}}} + \frac{0.75}{2} \right) \right] d \bar{\phi} + D_{4} \left(\frac{\phi}{\phi_{\mathrm{f}}}\right)^{3},
\end{eqnarray}
where $V_{\mathrm{f}}$ is a constant, $D_{4}/V_{\mathrm{f}}=1.5$ is fixed, and $C_{4}$ is a normalization constant with
\begin{eqnarray}
C_{4} = \int_{\phi_{\mathrm{f}}}^{0} \left[ \frac{\bar{\phi}}{\phi_{\mathrm{f}}} \left( \frac{\bar{\phi}}{\phi_{\mathrm{f}}} - 1 \right) \left( \frac{\bar{\phi}}{\phi_{\mathrm{f}}} + 0.75 \right) \left( \frac{\bar{\phi}}{\phi_{\mathrm{f}}} - \frac{1}{2} \right) \left( \frac{\bar{\phi}}{\phi_{\mathrm{f}}} + \frac{0.75}{2} \right) \right] d \bar{\phi}.
\end{eqnarray}
Note that in this case, if we choose a certain $S_{\mathrm{f}}$, the local maximum is shifted to $\simeq 0.9502 \times S_{\mathrm{f}}$, and we considered this effect.
\end{enumerate}
Therefore, for all cases, $V_{\mathrm{f}}$ and $S_{\mathrm{f}} = \sqrt{4\pi} \phi_{\mathrm{f}}$ (or $S_{\mathrm{t}} = \sqrt{4\pi} \phi_{\mathrm{t}}$) are free parameters of each potential.

\section{\label{sec:var}Various phenomena of bubble collisions}

\subsection{Bubble percolation}

First, we study bubble percolation. By percolation, we mean that two true vacuum bubbles are collided and have eventually emerged to one true vacuum region. The question is whether this process can continuously happen, and what is the new phenomena that could not be observed by the thin-wall approximation?

\subsubsection{Spherical symmetry}

\begin{figure}
\begin{center}
\includegraphics[scale=0.2]{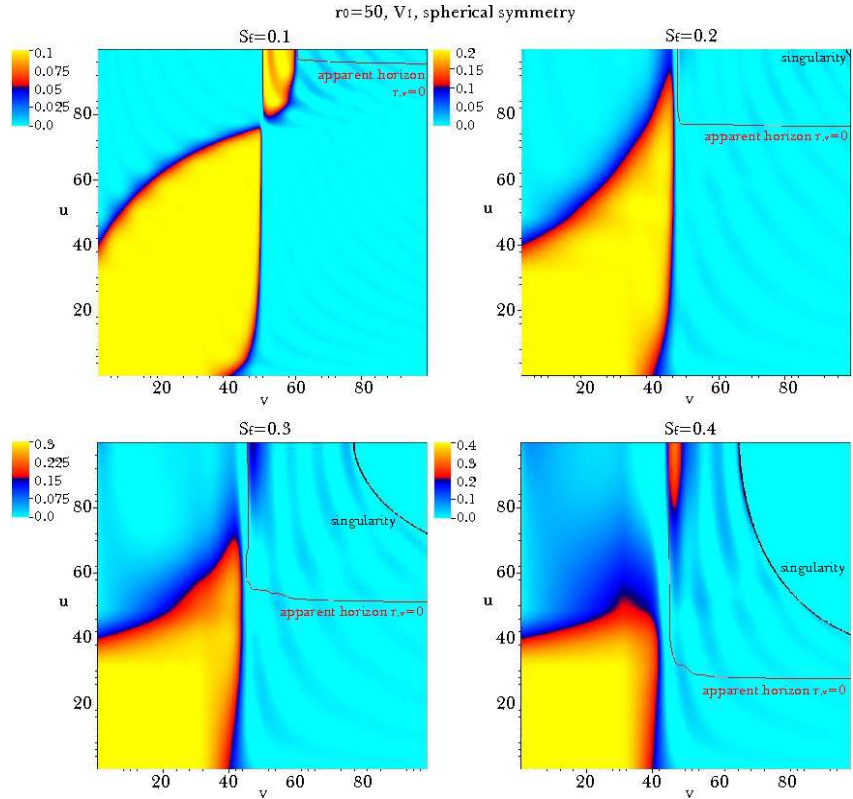}
\caption{\label{fig:per_sphe}Bubble percolation with spherical symmetry: $V_{1}$ with $V_{\mathrm{f}}=0.00001$, $r_{0}=50$, $\Delta u=\Delta v=20$, $u_{\mathrm{shell}}=v_{\mathrm{shell}}=30$, $S_{\mathrm{t}}=0$, and varying $S_{\mathrm{f}}=0.1, 0.2, 0.3, 0.4$.}
\end{center}
\end{figure}

Figure~\ref{fig:per_sphe} shows the collision of an in-going false vacuum bubble and an out-going true vacuum bubble. In this case, we choose the same field value for the true vacuum. Therefore, after the collision, bubbles should be emersed and be percolated. In these calculations, we used the potential $V_{1}$ with $V_{\mathrm{f}}=0.00001$. We also used $r_{0}=50$, $\Delta u=\Delta v=20$, $u_{\mathrm{shell}}=v_{\mathrm{shell}}=30$, $S_{\mathrm{t}}=0$, and varying $S_{\mathrm{f}}=0.1, 0.2, 0.3, 0.4$.

\begin{figure}
\begin{center}
\includegraphics[scale=0.4]{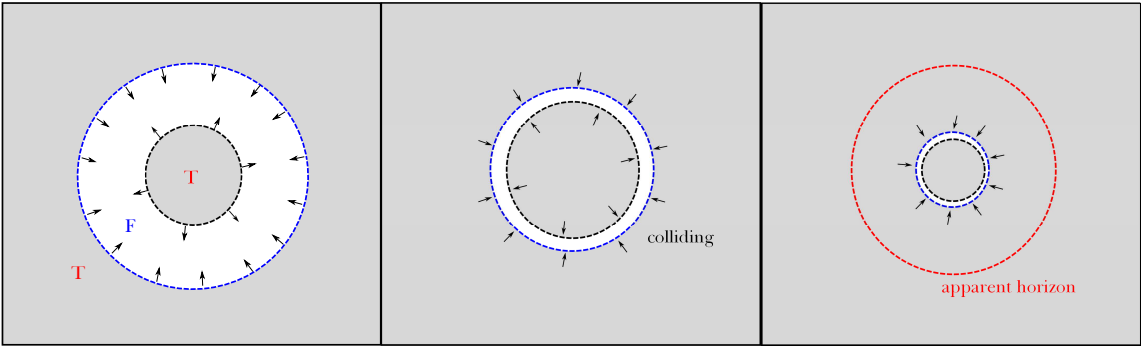}
\caption{\label{fig:spherical1}A schematic diagram of bubble percolation with the spherical symmetry.}
\end{center}
\end{figure}

As $S_{\mathrm{f}}$ increases, the tension and the total energy increase. Therefore, the location of the apparent horizon and singularity becomes lower and lower as $S_{\mathrm{f}}$ increases; this implies that the apparent horizon increases, and a singularity relatively quickly appears.

In Figure~\ref{fig:per_sphe}, the yellow region is the false vacuum region, while the skyblue region is the true vacuum region. The black colored region is around the local maximum of the potential $V_{1}$ and hence it shows the dynamics of the thick wall.

The behavior of the out-going true vacuum wall before colliding (for $v<30$) is changed as $S_{\mathrm{f}}$ varies. If $S_{\mathrm{f}}$ is sufficiently small, then initially it shrinks (upper left). However, as $S_{\mathrm{f}}$ increases, it tends to expand (lower right). Unless the shell has sufficient energy, it tends to collapse. In addition, if the shell has sufficient energy, it can expand, although the inner region becomes destabilized; the inside of the out-going shell ($v<30$ and $u>50$) is not exactly zero and the fields are slightly perturbed (we can compare the upper left and upper right diagrams). These results are consistent with that of the false vacuum bubble cases \cite{Hansen:2009kn}.

The upper left of Figure~\ref{fig:per_sphe} shows interesting behaviors: after the collision, some fields can roll up to the false vacuum region ($50<v<60$ and $u>80$). This is due to the fact that the colliding energy perturbs and pushes the field to the other local minimum. This is related to the vacuum destabilization by bubble collisions, and it will be discussed further in Section~\ref{sec:vacdeep}.

\begin{figure}
\begin{center}
\includegraphics[scale=0.6]{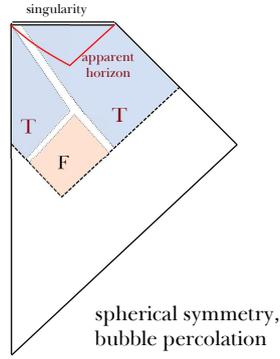}
\caption{\label{fig:causal1}Causal structure of percolating bubbles with spherical symmetry.}
\end{center}
\end{figure}

We can intuitively draw as in Figure~\ref{fig:spherical1}. Initially, there is an in-going false vacuum bubble and an out-going true vacuum bubble (left). The false vacuum region becomes narrower (middle). Eventually, the collided region collapses and shrinks beyond the apparent horizon (right). This can be compared to the Figure~\ref{fig:spherical2} in Section~\ref{sec:vacdeep}. This mimics the situation that bubble collisions can generate a black hole: Figure~\ref{fig:causal1}.

\begin{figure}
\begin{center}
\includegraphics[scale=0.2]{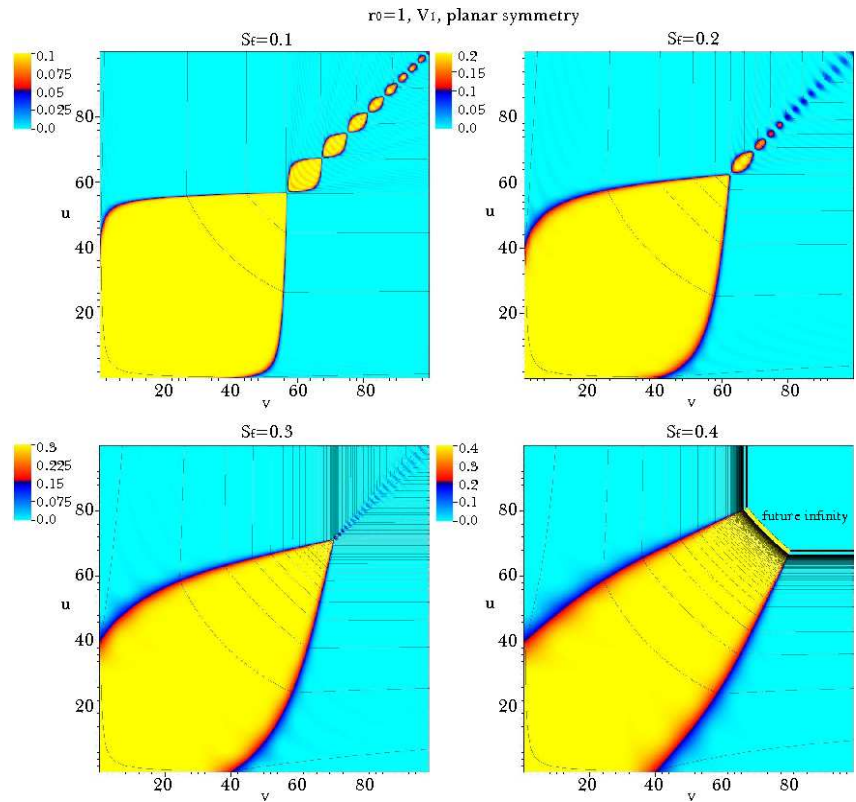}
\caption{\label{fig:per_plan}Bubble percolation with planar symmetry: $V_{1}$ with $V_{\mathrm{f}}=0.0001$, $r_{0}=1$, $\Delta u=\Delta v=20$, $u_{\mathrm{shell}}=v_{\mathrm{shell}}=30$, $S_{\mathrm{t}}=0$, and varying $S_{\mathrm{f}}=0.1, 0.2, 0.3, 0.4$. Black curves are contour curves of $r$ and the differences of each of the curves are $1$ and drawn from $r=0$ to $r=100$.}
\end{center}
\end{figure}

\subsubsection{Planar symmetry}

Figure~\ref{fig:per_plan} shows the bubble percolation of two planar bubbles. We assume these conditions: $V_{1}$ with $V_{\mathrm{f}}=0.0001$, $r_{0}=1$, $\Delta u=\Delta v=20$, $u_{\mathrm{shell}}=v_{\mathrm{shell}}=30$, $S_{\mathrm{t}}=0$, and varying $S_{\mathrm{f}}=0.1, 0.2, 0.3, 0.4$.

The first clear observation is that after the collision, walls oscillates back-and-forth (upper left). As a wall oscillates, oscillating field amplitudes decreases (upper right) and eventually the wall disappears (lower left).

When the tension of the shell is sufficiently small, the shell initially expands to the true vacuum region, although it eventually and quickly shrinks to the center (upper left). As the tension of the shell increases, the initial expansion of the shell becomes slower and slower (upper right and lower left). Eventually, if the shell energy is sufficiently large, it slowly approaches and it may not be able to collide, since the false vacuum region inflates too quickly (lower right).

We need more comments on the lower right of Figure~\ref{fig:per_plan}. For all diagrams in Figure~\ref{fig:per_plan}, we can see the details of the $r$ contours. In this case, $r$ is not a real radius; rather, this is just a metric function that relates the planar direction and the orthogonal direction. For the inside of the false vacuum region, it is space-like, while it is time-like for the true vacuum region. The $r$ function increases from lower left to the upper right; and it diverges in the lower right of Figure~\ref{fig:per_plan}. If we only see this diagram, it is not easy to interpret the space-like piece in this diagram, where the numerical simulation cannot work. It seems to not be a singularity, since two bubbles are not collided yet. To interpret correctly, we have to compare Figure~\ref{fig:per_hype_massless} in Section~\ref{sec:hypsym}; in the hyperbolic symmetry, we can interpret that the region is future infinity, where the time-like parameter diverges. Then we can finally interpret the figure that two bubbles cannot collide since the false vacuum region is rapidly expanded.

\subsubsection{\label{sec:hypsym}Hyperbolic symmetry}

\begin{figure}
\begin{center}
\includegraphics[scale=0.2]{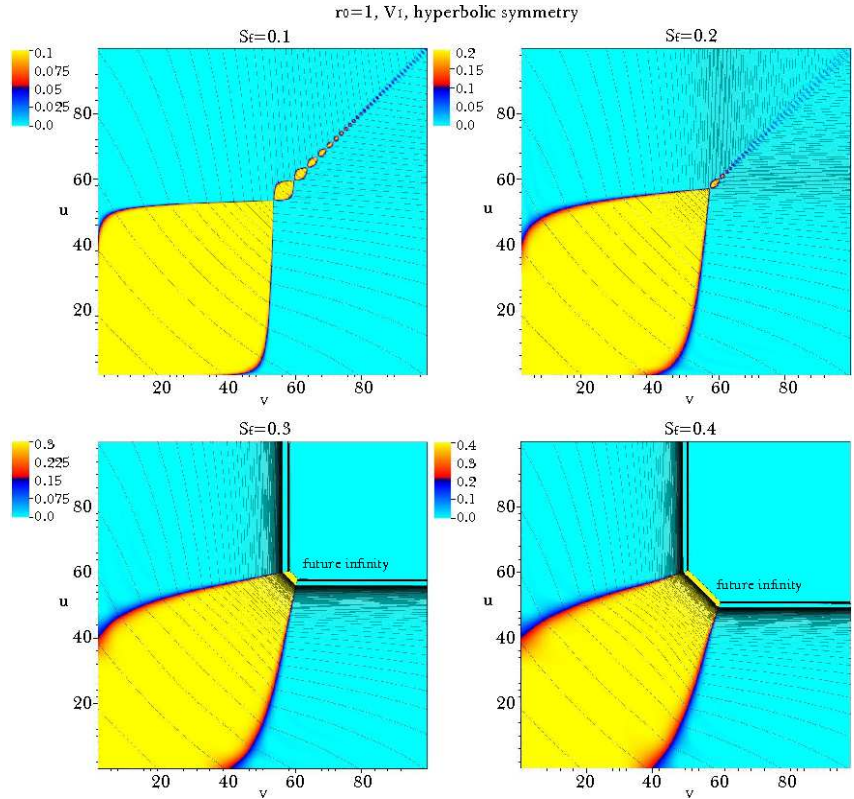}
\caption{\label{fig:per_hype_massless}Bubble percolation with hyperbolic symmetry: $V_{1}$ with $V_{\mathrm{f}}=0.0001$, $r_{0}=1$, $\Delta u=\Delta v=20$, $u_{\mathrm{shell}}=v_{\mathrm{shell}}=30$, $S_{\mathrm{t}}=0$, and varying $S_{\mathrm{f}}=0.1, 0.2, 0.3, 0.4$. Black curves are contour curves of $r$ and the differences of each of the curves are $10$ and drawn from $r=0$ to $r=1000$.}
\end{center}
\end{figure}

\paragraph{Massless case} In Figure~\ref{fig:per_hype_massless}, we observe bubble percolation with the hyperbolic symmetry. The results are similar to the planar symmetric case. We used $V_{1}$ with $V_{\mathrm{f}}=0.0001$, $r_{0}=1$, $\Delta u=\Delta v=20$, $u_{\mathrm{shell}}=v_{\mathrm{shell}}=30$, $S_{\mathrm{t}}=0$, and varying $S_{\mathrm{f}}=0.1, 0.2, 0.3, 0.4$.

When two bubbles collide, the collided region oscillates back-and-forth (upper left). The collided region eventually decays and disappears (upper right). Similar to the planar symmetric case, as the tension increases, the wall slowly moves. For all cases, $r$ contours are always space-like. This is not strange, since our metric ansatz assumed $r$ as a time-like parameter. Therefore, $r$ works as a global time function. For figures in the lower left and lower right in Figure~\ref{fig:per_hype_massless}, the function $r$ diverges as it approaches the upper right region of the figure. Therefore, we can interpret that two bubbles may not be able to collide and the space-like boundary is the future infinity of the false vacuum region, where the in-going and out-going null boundaries are future infinity of the Minkowski regions.

\begin{figure}
\begin{center}
\includegraphics[scale=0.6]{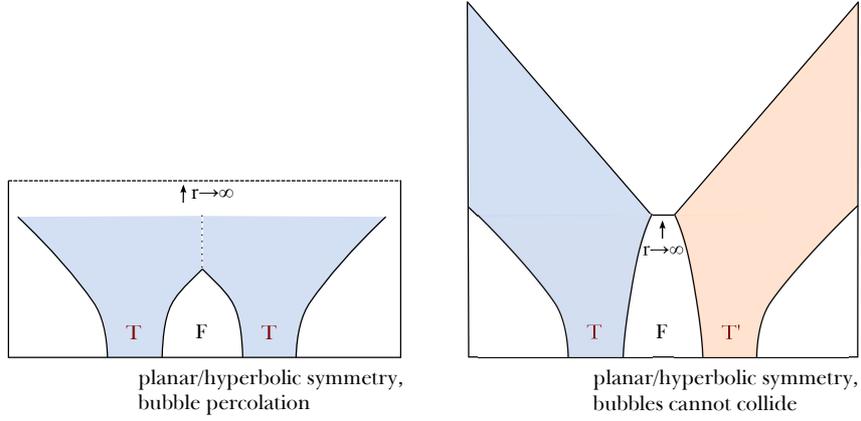}
\caption{\label{fig:causal2}Causal structure of percolating bubbles with planar/hyperbolic symmetry. Left: bubbles are percolated. Right: when the tension increases, bubbles may not be able to collide.}
\end{center}
\end{figure}

To summarize, we obtain two interesting causal structures in Figure~\ref{fig:causal2}: left is normal bubble percolation when the tension is sufficiently small; right is the large tension case and due to the repulsiveness of large tension walls, bubbles cannot collide and bubble walls touch the future infinity.


\begin{figure}
\begin{center}
\includegraphics[scale=0.2]{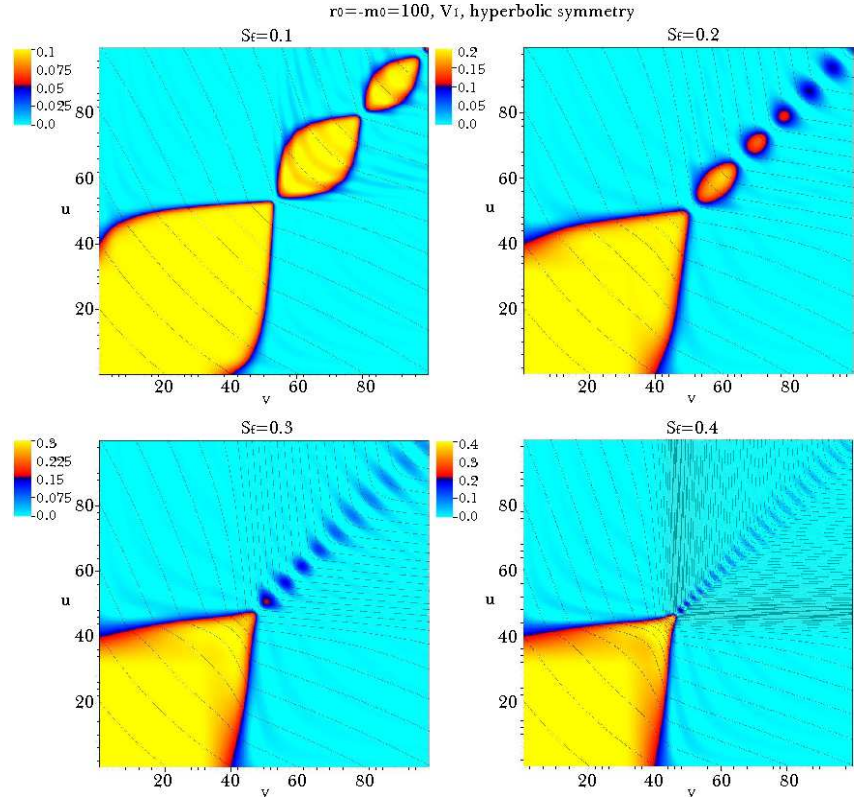}
\caption{\label{fig:per_hype}Bubble percolation with hyperbolic symmetry: $V_{1}$ with $V_{\mathrm{f}}=0.0001$, $r_{0}=-m_{0}=100$, $\Delta u=\Delta v=20$, $u_{\mathrm{shell}}=v_{\mathrm{shell}}=30$, $S_{\mathrm{t}}=0$, and varying $S_{\mathrm{f}}=0.1, 0.2, 0.3, 0.4$. Black curves are contour curves of $r$ and the differences of each of the curves are $10$ and drawn from $r=100$ to $r=1100$.}
\end{center}
\end{figure}

\paragraph{Massive case} We can also choose the $m_{0} < 0$ case. In Figure~\ref{fig:per_hype}, we especially choose $-m_{0} = r_{0} = 100$, since $m_{0}$ and $r_{0}$ should be the same order to see the effect of the mass term. We used $V_{1}$ with $V_{\mathrm{f}}=0.0001$, $\Delta u=\Delta v=20$, $u_{\mathrm{shell}}=v_{\mathrm{shell}}=30$, $S_{\mathrm{t}}=0$, and varying $S_{\mathrm{f}}=0.1, 0.2, 0.3, 0.4$.

As $S_{\mathrm{f}}$ increases, it is qualitatively similar to the massless case; however, the wall collides relatively rapidly than the massless case. Therefore, the existence of mass allows a wide range of tensions that allows collision. Therefore, for convenience, we choose this kind of initial conditions for the latter sections, since we want to build colliding bubbles.

\begin{figure}
\begin{center}
\includegraphics[scale=0.2]{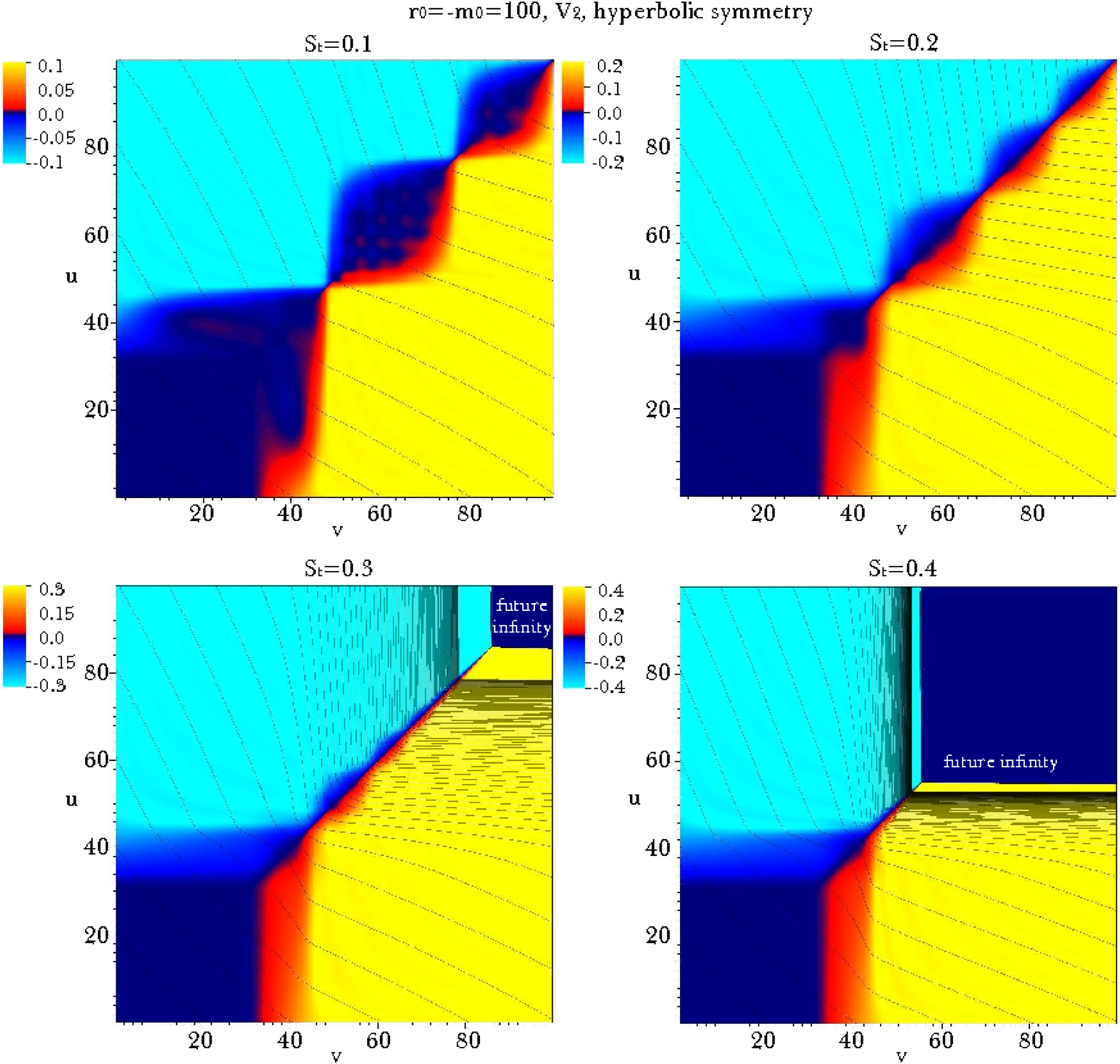}
\caption{\label{fig:fie_hype}Bubble collisions of symmetric different true vacua with hyperbolic symmetry: $V_{2}$ with $V_{\mathrm{f}}=0.0001$, $r_{0}=-m_{0}=100$, $\Delta u=\Delta v=20$, $u_{\mathrm{shell}}=v_{\mathrm{shell}}=30$, $S_{\mathrm{f}}=0$, and varying $S_{\mathrm{t}}=0.1, 0.2, 0.3, 0.4$. Black curves are contour curves of $r$ and the differences of each of the curves are $10$ and drawn from $r=100$ to $r=1100$.}
\includegraphics[scale=0.6]{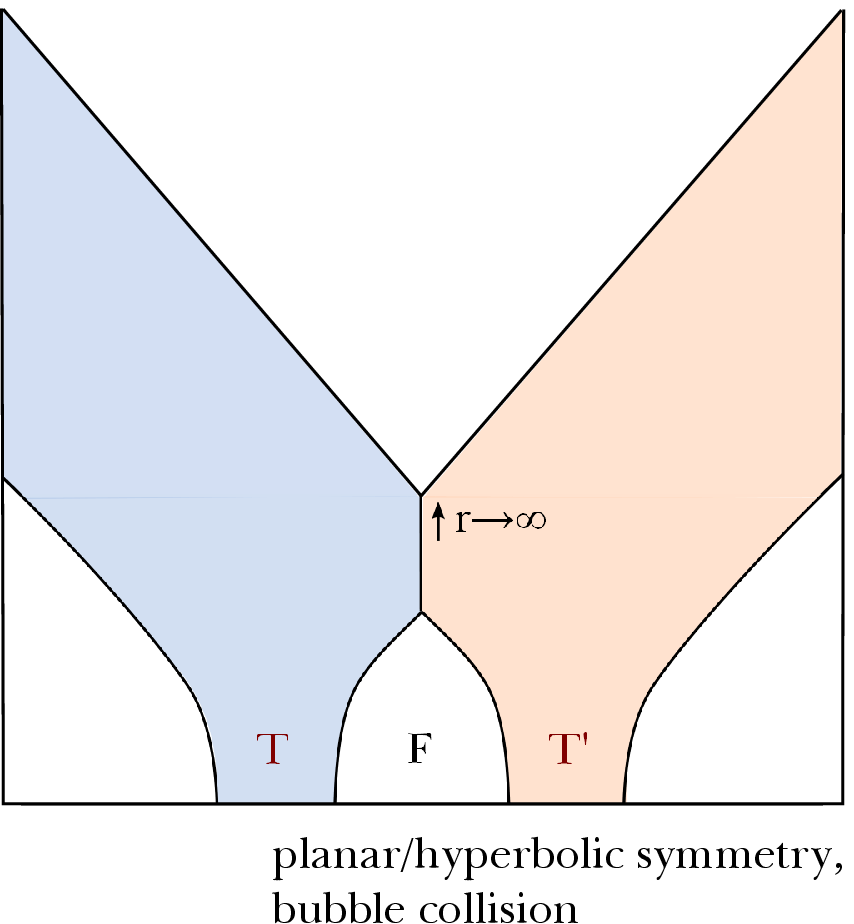}
\caption{\label{fig:causal3}Causal structure of colliding symmetric bubbles with different field values.}
\end{center}
\end{figure}

\subsection{Dynamics of bubble walls}

Second, we study dynamics of bubble walls. In this case, we especially focus on the case that the left true vacuum region and the right true vacuum region have different field values. Therefore, the bubble wall region will not disappear. In this section, we will observe their back-reaction to the causal structure. Also, we see when the initial conditions of the left and the right are asymmetric.

\subsubsection{Symmetric bubbles}

In Figure~\ref{fig:fie_hype}, we studied bubble collisions of symmetric bubbles with different field values of true vacuum region: $V_{2}$ with $V_{\mathrm{f}}=0.0001$, $r_{0}=-m_{0}=100$, $\Delta u=\Delta v=20$, $u_{\mathrm{shell}}=v_{\mathrm{shell}}=30$, $S_{\mathrm{f}}=0$, and varying $S_{\mathrm{t}}=0.1, 0.2, 0.3, 0.4$. Note that we only used hyperbolic symmetry, since the planar case is qualitatively similar to the hyperbolic case.

As $S_{\mathrm{t}}$ increases, the oscillating frequency of the thick wall increases and is quickly stabilized for sufficiently large $S_{\mathrm{t}}$. One interesting thing to notice is that after a sufficient time, two regions cannot be emersed, since two regions have different field values. Therefore, between two true vacuum regions, there should be always a false vacuum region (blue colored region). In a finite advanced and retarded time, the function $r$ diverges. Therefore, we can interpret that the coordinate time diverges and hence there is the future infinity. For the true vacuum part, the future infinities are null. Between two future null infinities, the false vacuum region can roll a crossing point between two null boundaries. This is a quite different behavior than the bubble percolation case (Figure~\ref{fig:causal3}).

\subsubsection{Asymmetric bubbles}

\begin{figure}
\begin{center}
\includegraphics[scale=0.2]{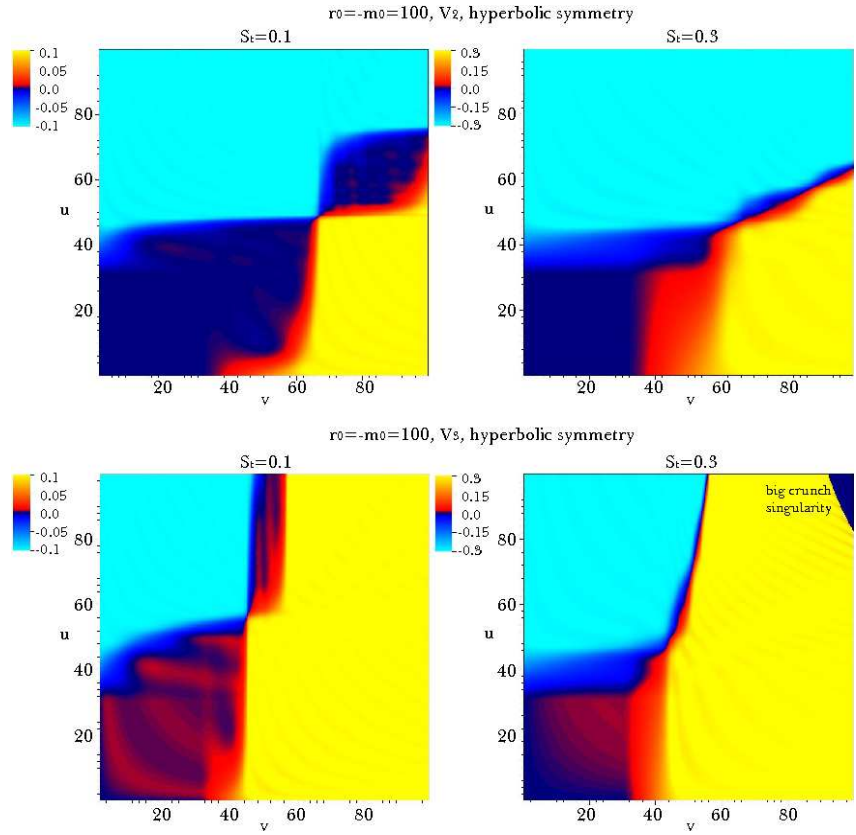}
\caption{\label{fig:asy_hype}Upper: Bubble collisions of asymmetric different true vacua with hyperbolic symmetry: $V_{2}$ with $V_{\mathrm{f}}=0.0001$, $r_{0}=-m_{0}=100$, $\Delta u=20$, $\Delta v=40$, $u_{\mathrm{shell}}=v_{\mathrm{shell}}=30$, $S_{\mathrm{f}}=0$, and varying $S_{\mathrm{t}}=0.1, 0.3$. Lower: Bubble collisions of asymmetric different true vacua with hyperbolic symmetry: $V_{3}$ with $V_{\mathrm{f}}=0.0001$, $r_{0}=-m_{0}=100$, $\Delta u=\Delta v=20$, $u_{\mathrm{shell}}=v_{\mathrm{shell}}=30$, $S_{\mathrm{f}}=0$, and varying $S_{\mathrm{t}}=0.1, 0.3$.}
\end{center}
\end{figure}

In Figure~\ref{fig:asy_hype}, we studied bubble collisions of asymmetric bubbles with different field values of true vacuum region. First we change tensions of two walls: $V_{2}$ with $V_{\mathrm{f}}=0.0001$, $r_{0}=-m_{0}=100$, $\Delta u=20$, $\Delta v=40$, $u_{\mathrm{shell}}=v_{\mathrm{shell}}=30$, $S_{\mathrm{f}}=0$, and varying $S_{\mathrm{t}}=0.1, 0.3$. Second we change vacuum energy between two true vacuum regions: $V_{3}$ with $V_{\mathrm{f}}=0.0001$, $r_{0}=100$, $\Delta u=\Delta v=20$, $u_{\mathrm{shell}}=v_{\mathrm{shell}}=30$, $S_{\mathrm{f}}=0$, and varying $S_{\mathrm{t}}=0.1, 0.3$.

\paragraph{Asymmetry of tension}

If one side has larger tension (in our case, the skyblue side has larger tension, since the thickness of the shell is narrower than the yellow side), it can push the smaller tension region, since it is related to the energy of the shell. The upper two figures show this behavior.

\paragraph{Asymmetry of vacuum energy}

If one true vacuum side has smaller vacuum energy (in our case, the yellow region is anti de Sitter and hence it is smaller than the skyblue region), it can push the larger vacuum energy region, since the difference of the vacuum energy between the false vacuum and the true vacuum is changed to the pressure of the wall. Therefore, in the lower figures, the yellow region pushes the skyblue region.

The lower right in Figure~\ref{fig:asy_hype} is important to emphasize. This shows the collision between a de Sitter vacuum and an anti de Sitter vacuum. This problem was systematically discussed by \cite{Freivogel:2007fx}. However, here we can see details from numerical calculations.

\begin{figure}
\begin{center}
\includegraphics[scale=0.3]{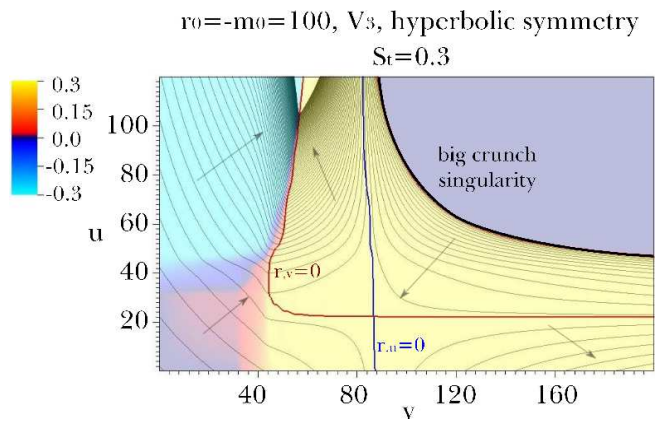}
\caption{\label{fig:asy_hype_wider}Bubble collisions of asymmetric different true vacua with hyperbolic symmetry: $V_{3}$ with $V_{\mathrm{f}}=0.0001$, $r_{0}=-m_{0}=100$, $\Delta u=\Delta v=20$, $u_{\mathrm{shell}}=v_{\mathrm{shell}}=30$, $S_{\mathrm{f}}=0$, and $S_{\mathrm{t}}=0.3$. Black contours are $r$ and the differences of each of the contours are $10$ and drawn from $r=0$ to $r=1000$. Red and blue curves are $r_{,v}=0$ and $r_{,u}=0$ contours. Black arrows are directions where $r$ increases.}
\end{center}
\end{figure}
\begin{figure}
\begin{center}
\includegraphics[scale=1]{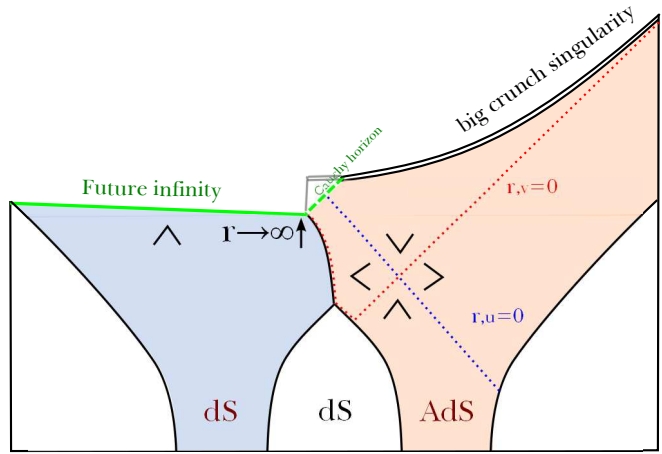}
\caption{\label{fig:asy}Causal structure of bubble collisions between de Sitter and anti de Sitter bubbles. We denote local horizons and Bousso wedges.}
\end{center}
\end{figure}

Figure~\ref{fig:asy_hype_wider} is the detailed version of the lower right of Figure~\ref{fig:asy_hype}. Black contours are $r$ and the distances of each of the contours are $10$. Red and blue curves are $r_{,v}=0$ and $r_{,u}=0$ contours. Black arrows are directions where $r$ increases. Therefore, for the de Sitter region, $r$ contours are always space-like. However, inside of the anti de Sitter region, the signature of $r$ changed. After $r$ changes the signature two times ($r_{,u}<0$ and $r_{,v}<0$), eventually  $r$ decreases to zero. This whole smooth evolution of the bubble collision and the formation of big crunch singularity are possible due to the numerical calculations.

We summarized in Figure~\ref{fig:asy}. This is the advanced version of the previous work \cite{Freivogel:2007fx}. The left bubble is de Sitter (in principle, Minkowski is possible). The left bubble is collided into the right anti de Sitter bubble. The wall moves in a time-like direction, is biased to the left direction, and eventually becomes parallel to the null direction; the tension via the difference of vacuum energy is larger for the anti de Sitter side and the speed of the wall will eventually approach the speed of light. The future infinity of the left side is space-like (for Minkowski, it should be null). There is a future infinity on the wall, and the wall will touch the future infinity of the left side. When the wall touches the future infinity of the left side, the wall may or may not generate a Cauchy horizon; this depends on initial conditions. However, at least in the data of Figure~\ref{fig:asy_hype_wider}, we can see that \textit{there should be a Cauchy horizon} in the anti de Sitter side (green dotted line) because of two reasons: (1) the wall almost moves in the left-going null direction and (2) the right side of the wall is approximately pure anti de Sitter as time goes on, and therefore, the blue curve ($r_{,u}=0$ horizon) should be parallel to the null direction. Therefore, the wall and the $r_{,u}=0$ horizon will not meet each other, and hence, there should be a Cauchy horizon for the anti de Sitter side. We also notice the Bousso wedges, that implies the direction where $r$ increases. Due to the horizons, the Bousso wedges are changed.

There are two main differences between the thin-wall approximation \cite{Freivogel:2007fx} and our numerical studies:
\begin{enumerate}
\item The crossing point of two horizons ($r_{,v}=0$ and $r_{,u}=0$) can happen before the causal future of the bubble collision, while this was not allowed by the thin-wall approximation. This is not strange, because the horizons are dynamically deformed \textit{on the wall} and connected between de Sitter and anti de Sitter regions.
\item The future infinity can have the \textit{Cauchy horizon} as we have commented. However, the time parameter $r$ will diverge on the wall when the wall touches the future boundary. Therefore, in terms of the analytic coordinates of the anti de Sitter side, they should be identified to the time-like $r=\infty$ boundary (gray diagrams beyond the Cauchy horizon in Figure~\ref{fig:asy}), although we cannot determine beyond the Cauchy horizon from the past data.
\end{enumerate}

\subsection{\label{sec:vacdeep}Vacuum destabilization to deeper vacuum}


Finally, we focus on the vacuum destabilization to deeper vacuum by bubble collisions. We tested two cases. First, for spherical symmetry, $V_{4}$ with $V_{\mathrm{f}}=0.00001$, $r_{0}=50$, $\Delta u=\Delta v=20$, $u_{\mathrm{shell}}=v_{\mathrm{shell}}=30$, $S_{\mathrm{t}}=0$, and $S_{\mathrm{f}}=0.1$. Second, for hyperbolic symmetry, $V_{4}$ with $V_{\mathrm{f}}=0.0001$, $r_{0}=-m_{0}=100$, $\Delta u=\Delta v=20$, $u_{\mathrm{shell}}=v_{\mathrm{shell}}=30$, $S_{\mathrm{t}}=0$, and $S_{\mathrm{f}}=0.1$.

\begin{figure}
\begin{center}
\includegraphics[scale=0.2]{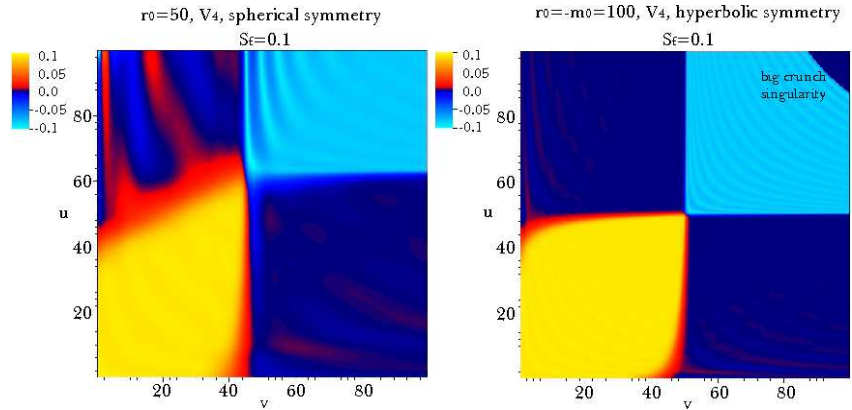}
\caption{\label{fig:dee}Transition to deeper vacuum. Upper: Spherical symmetry: $V_{4}$ with $V_{\mathrm{f}}=0.00001$, $r_{0}=50$, $\Delta u=\Delta v=20$, $u_{\mathrm{shell}}=v_{\mathrm{shell}}=30$, $S_{\mathrm{t}}=0$, and $S_{\mathrm{f}}=0.1$. Lower: Bubble collisions of asymmetric different true vacua with hyperbolic symmetry: $V_{4}$ with $V_{\mathrm{f}}=0.0001$, $r_{0}=-m_{0}=100$, $\Delta u=\Delta v=20$, $u_{\mathrm{shell}}=v_{\mathrm{shell}}=30$, $S_{\mathrm{t}}=0$, and $S_{\mathrm{f}}=0.1$.}
\end{center}
\end{figure}
\begin{figure}
\begin{center}
\includegraphics[scale=0.4]{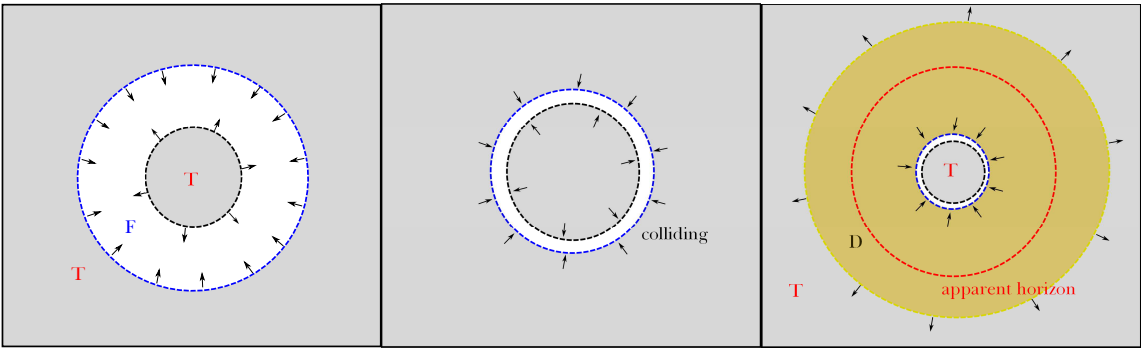}
\caption{\label{fig:spherical2}Deeper vacuum transition by bubble collisions, for spherical symmetric case.}
\end{center}
\end{figure}
\begin{figure}
\begin{center}
\includegraphics[scale=0.6]{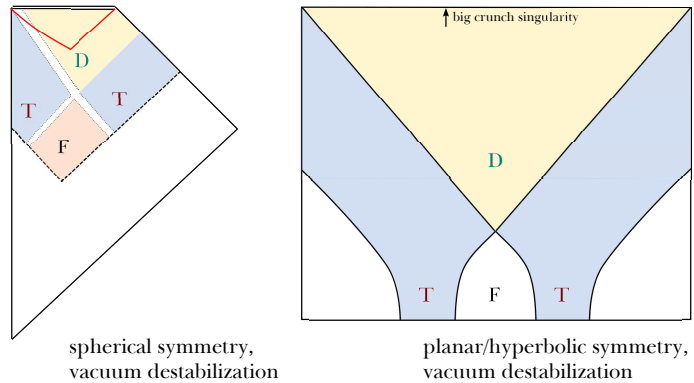}
\caption{\label{fig:causal4}Causal structure of deeper vacuum transitions via bubble collisions with spherical (left) and hyperbolic (right) symmetry.}
\end{center}
\end{figure}

Figure~\ref{fig:dee} shows the spherical and hyperbolic cases. The skyblue region is the deeper vacuum.

For the spherical case (left of Figure~\ref{fig:dee}), after two walls are collided, a deeper region is generated. The boundary between the deeper vacuum and the true vacuum moves along the out-going null direction. Hence, it can be outside of the event horizon, if we properly choose the tension. This can be interpreted by Figure~\ref{fig:spherical2}. After the bubble collision, the deeper vacuum region expanded along the out-going direction, and hence it can be outside of the apparent horizon.

For the hyperbolic case (right), the deeper vacuum region and the big crunch singularity of the anti de Sitter region are symmetrically generated, as one expected. To summarize, we draw the causal structures for vacuum destabilization for spherical and hyperbolic cases in Figure~\ref{fig:causal4}.

\section{\label{sec:dis}Discussion}

In this paper, we studied bubble collisions with gravitation by numerical calculations. We used the double-null formalism to implement numerical simulations. We used three types of symmetries: spherical, planar, and hyperbolic. Three kinds of symmetries can be qualitatively useful in various situations during the first order phase transition (Figure~\ref{fig:bubbles}). In this discussion, we will summarize our results.

First, we tested bubble percolation by bubble collisions. After bubbles collided, the false vacuum region disappeared and two true vacuum regions are emersed. For the spherical symmetric case, it will eventually induce a black hole in a true vacuum background (Figure~\ref{fig:causal1}). For the planar or hyperbolic symmetric case, two bubbles are emersed (left of Figure~\ref{fig:causal2}). However, as tension increases, if the wall is not sufficient to push the false vacuum region, then two walls may not be able to meet (right of Figure~\ref{fig:causal2}).

Second, we tested bubble collisions with different field values to see the dynamics of bubble walls. For symmetric cases, the collided wall moves symmetrically and the wall will touch the future infinity between two true vacuum regions (Figure~\ref{fig:causal3}). For asymmetric cases, the wall is biased due to the difference of pressures. The interesting example is the collision of the de Sitter vacuum and the anti de Sitter vacuum, and the causal structure is analyzed in Figure~\ref{fig:asy}.

Finally, we tested vacuum destabilization due to the bubble collisions. For the spherical symmetric case, the deeper vacuum region can be outside of the event horizon (left of Figure~\ref{fig:causal4}). When we compare this to Figure~\ref{fig:causal1}, we can notice that this can give a significant difference for outside observers. For the hyperbolic symmetric case, the deeper vacuum region is expanded over the null directions and there will be a space-like big crunch singularity (right of Figure~\ref{fig:causal4}).

There were a number of previous studies on bubble collisions, and this paper can confirm these studies. In addition, there are mainly new contributions provided by this paper.
\begin{itemize}
\item The spherical symmetric bubble collision (including black hole formation) was studied numerically with gravitation (Figure~\ref{fig:per_sphe} and left of Figure~\ref{fig:dee}). This can give a \textit{general intuition for multiple bubble collisions} (Figures~\ref{fig:spherical1} and \ref{fig:spherical2}).\footnote{Of course, we do not claim that all multiple bubble collisions generate a black hole; rather, a formation of a black hole requires a concentration of energy and our study shows that a black hole can be formed in these special cases. See the references in \cite{Khlopov}.}
\item We observed more general initial conditions than Coleman-DeLuccia type bubbles. This was easily done, because we used the double-null formalism. One interesting observation is that as tension increases \textit{walls slowly moves} and they may not be collided (lower right in Figure~\ref{fig:per_plan} and lower left and right in Figure~\ref{fig:per_hype_massless}).
\item The smooth transitions of metric functions (especially $r$) are realized. This cannot be done by the thin-wall approximation. We obtained future boundary formation from smooth initial data without assuming thin-wall and classified various causal structures. Moreover, Figure~\ref{fig:asy} shows more complicated behavior of horizons of $r$ during bubble collisions.
\end{itemize}

There can be further applications and there will be many other models and issues for bubble collisions. Also, there can be many problems that can be solved by double-null formalism with hyperbolic or planar symmetry. We leave these for future work.

\section*{Acknowledgment}

The authors would like to thank Ewan Stewart for discussions and encouragement. The authors also would like to thank Maxim Khlopov, Matthew Kleban and Dong-Hoon Kim for useful comments on this work and other references. DY, BHL, and WL were supported by the National Research Foundation of Korea grant funded by the Korean government (MEST) through the Center for Quantum Spacetime (CQUeST) of Sogang University with grant number 2005-0049409. DH was supported by Korea Research Foundation grants (KRF-313-2007-C00164, KRF-341-2007-C00010) funded by the Korean government (MOEHRD) and BK21. WL was supported by the National Research Foundation of Korea grant funded by the Korean government (MEST) [NRF-2010-355-C00017].

\begin{figure}
\begin{center}
\includegraphics[scale=1]{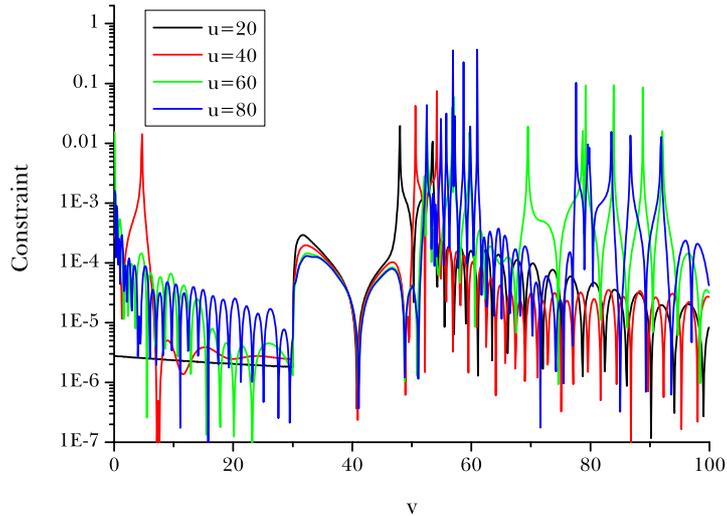}
\caption{\label{fig:consistency}Consistency test for Equation~(\ref{eq:constraint}) along $u=20, 40, 60, 80$.}
\end{center}
\end{figure}
\begin{figure}
\begin{center}
\includegraphics[scale=1]{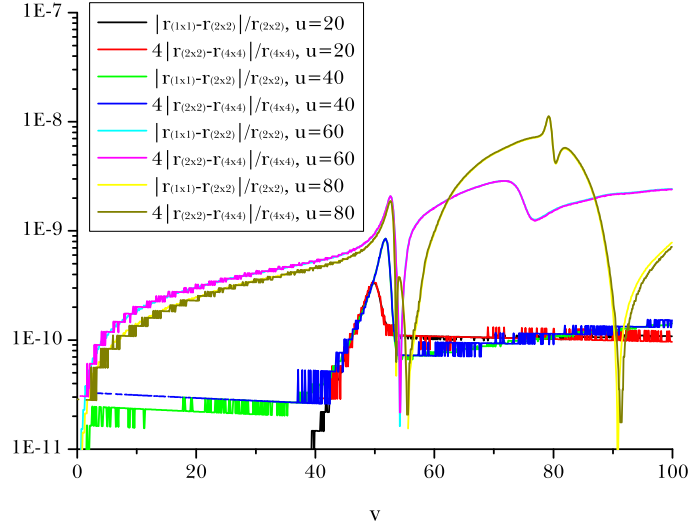}
\caption{\label{fig:convergence}Convergence test: $|r_{(1\times1)}-r_{(2\times2)}|/r_{(2\times2)}$ and $4|r_{(2\times2)}-r_{(4\times4)}|/r_{(4\times4)}$ along $u=20, 40, 60, 80$. This shows the second order convergence.}
\end{center}
\end{figure}

\section*{Appendix A: Numerical scheme}

The integration scheme that we used is the second order Runge-Kutta method. Here, we briefly discuss the integration scheme. As we see in Section~\ref{sec:imp}, we organized all equations by a set of first order differential equations. Let us assume that we know all data for initial surfaces $u=u_{\textit{i}}$ and $v=v_{\textit{i}}$ as we discussed in Section~\ref{sec:ini}, and observe how we can determine the point $(u=\Delta u + u_{\textit{i}}, v=\Delta v + v_{\textit{i}})$. For example, to obtain the function $r$, we have to solve
\begin{eqnarray}
r_{,u} &=& f,\label{eq:example1}\\
f_{,u} &=& \mathrm{Right\; hand\; side\; of\; Equation~(\ref{eq:E1})}\label{eq:example2}.
\end{eqnarray}
Of course, this is not the unique choice and we can use
\begin{eqnarray}
r_{,v} &=& g,\label{eq:example3}\\
g_{,v} &=& \mathrm{Right\; hand\; side\; of\; Equation~(\ref{eq:E2})},\label{eq:example4}
\end{eqnarray}
or
\begin{eqnarray}
r_{,v} &=& g, \label{eq:example5}\\
g_{,u} &=& \mathrm{Right\; hand\; side\; of\; Equation~(\ref{eq:E3})},\label{eq:example6}
\end{eqnarray}
and so on. In any case, all equations are first order differential equations. Therefore, for example, if we apply Equations~(\ref{eq:example1}) and (\ref{eq:example2}), we can successively solve
\begin{eqnarray}
f(\Delta u + u_{\textit{i}}, \Delta v + v_{\textit{i}}) &=& f(u_{\textit{i}}, \Delta v + v_{\textit{i}}) + \left( \mathrm{RHS\; of\; Equation~(\ref{eq:E1})\; at\;} (u_{\textit{i}}, \Delta v + v_{\textit{i}}) \right) \times \Delta u, \\
r(\Delta u + u_{\textit{i}}, \Delta v + v_{\textit{i}}) &=& r(u_{\textit{i}}, \Delta v + v_{\textit{i}}) + f(u_{\textit{i}}, \Delta v + v_{\textit{i}}) \times \Delta u.
\end{eqnarray}
Or, if we use Equations~(\ref{eq:example3}) and (\ref{eq:example4}), then we can successively solve
\begin{eqnarray}
g(\Delta u + u_{\textit{i}}, \Delta v + v_{\textit{i}}) &=& g(\Delta u + u_{\textit{i}}, v_{\textit{i}}) + \left( \mathrm{RHS\; of\; Equation~(\ref{eq:E2})\; at\;} (\Delta u + u_{\textit{i}}, v_{\textit{i}}) \right) \times \Delta v, \\
r(\Delta u + u_{\textit{i}}, \Delta v + v_{\textit{i}}) &=& r(\Delta u + u_{\textit{i}}, v_{\textit{i}}) + g(\Delta u + u_{\textit{i}}, v_{\textit{i}}) \times \Delta v.
\end{eqnarray}
If we use Equations~(\ref{eq:example5}) and (\ref{eq:example6}), then we can successively solve
\begin{eqnarray}
g(\Delta u + u_{\textit{i}}, \Delta v + v_{\textit{i}}) &=& g(u_{\textit{i}}, \Delta v + v_{\textit{i}}) + \left( \mathrm{RHS\; of\; Equation~(\ref{eq:E3})\; at\;} (u_{\textit{i}}, \Delta v + v_{\textit{i}}) \right) \times \Delta u, \\
r(\Delta u + u_{\textit{i}}, \Delta v + v_{\textit{i}}) &=& r(\Delta u + u_{\textit{i}}, v_{\textit{i}}) + g(\Delta u + u_{\textit{i}}, v_{\textit{i}}) \times \Delta v.
\end{eqnarray}
Although the actual calculations are more difficult than this since many functions are involved, the principle is the same. Note that these are based on the first order method (the Euler method). We can improve the integration scheme to the second order \cite{nr}: to solve the equation
\begin{eqnarray}
\frac{dy}{dx} = F(x,y),
\end{eqnarray}
we successively solve
\begin{eqnarray}
k_{1} &=& \Delta x \times F(x_{n},y_{n}),\\
k_{2} &=& \Delta x \times F(x_{n}+\Delta x / 2,y_{n}+k_{1}/2),\\
y_{n+1} &=& y_{n} + k_{2} + \mathcal{O}((\Delta x)^{3}),
\end{eqnarray}
where $n$ is the step and $\Delta x$ is the step size. It is not difficult to implement the second order Runge-Kutta method to the series of equations of this paper.

\section*{Appendix B: Consistency and convergence tests}

In this appendix, we report on the convergence and consistency tests for our simulations. As a demonstration, we check the case of hyperbolic symmetry: $V_{1}$ with $V_{\mathrm{f}}=0.0001$, $r_{0}=-m_{0}=100$, $\Delta u=\Delta v=20$, $u_{\mathrm{shell}}=v_{\mathrm{shell}}=30$, $S_{\mathrm{t}}=0$, and $S_{\mathrm{f}}=0.1$.

For consistency, we test one of the constraint functions:
\begin{equation}\label{eq:constraint}
\frac{\left|g_{,v}-2gd+4\pi r T^{\phi}_{vv}\right|}{\left(\left|g_{,v}\right|+\left|2gd - 4\pi r T^{\phi}_{vv}\right|\right)/2}
\end{equation}
around $u=20, 40, 60, 80$. Figure~\ref{fig:consistency} shows that it is less than $1$~\% except some points, where the denominator oscillatory vanishes ($g_{,v}\approx 0$); this will not be accumulated as one integrates along $v$. Therefore, this shows good consistency.

For convergence, we compared finer simulations: $1\times1$, $2\times2$, and $4\times4$ times finer for around $u=20, 40, 60, 80$.
In Figure~\ref{fig:convergence}, we see that the difference between the $1\times1$ and $2\times2$ times finer cases is $4$ times the difference between the $2\times2$ and $4\times4$ times finer cases,
and thus our simulation converges to second order. The numerical error is $\lesssim 10^{-6}\%$.

\end{document}